# What are Hybrid Development Methods Made Of? An Evidence-based Characterization

Paolo Tell[1], Jil Klünder[2], Steffen Küpper[3], David Raffo[4], Stephen G. MacDonell[5], Jürgen Münch[6], Dietmar Pfahl[7], Oliver Linssen[8], Marco Kuhrmann[9]

*[1]IT University Copenhagen, Denmark, Email: pate@itu.dk;*
*[2]Leibniz University Hannover, Germany, Email: jil.kluender@inf.uni-hannover.de;*
*[3,9]Clausthal University of Technology, Germany, Email: {steffen.kuepper, marco.kuhrmann}@tu-clausthal.de;*
*[4]Portland State University, USA, Email: raffod@pdx.edu;*
*[5]Auckland University of Technology, Auckland, New Zealand, Email: stephen.macdonell@aut.ac.nz;*
*[6]Reutlingen University, Germany, Email: juergen.muench@reutlingen-university.de;*
*[7]University of Tartu, Estonia, Email: dietmar.pfahl@ut.ee;*
*[8]FOM Hochschule für Oekonomie & Management, Germany, Email: oliver.linssen@fom.de*

## Abstract

*Among the multitude of software development pro- cesses available, hardly any is used by the book. Regardless of company size or industry sector, a majority of project teams and companies use customized processes that combine different development methods—so-called hybrid development methods. Even though such hybrid development methods are highly individualized, a common understanding of how to systematically construct synergetic practices is missing. In this paper, we make a first step towards devising such guidelines. Grounded in 1,467 data points from a large-scale online survey among practitioners, we study the current state of practice in process use to answer the question: What are hybrid development methods made of? Our findings reveal that only eight methods and few practices build the core of modern software development. This small set allows for statistically constructing hybrid development methods. Using an 85% agreement level in the participants' selections, we provide two examples illustrating how hybrid development methods are characterized by the practices they are made of. Our evidence- based analysis approach lays the foundation for devising hybrid development methods.*

## 1. INTRODUCTION

Today, companies often use highly individualized processes to run projects, often by integrating agile methods with their processes. For instance, Dikert et al. [1] found choosing and customizing an agile model to be an important success factor, and that agility in general changed the way software is developed. Dingsøyr et al. [2] reflect on a decade of agile methodologies and there is no denial that agile methods have become an important asset in many companies' process portfolios [3]–[6]. However, agile methods are not implemented by the book [7], [8], and in 2011, West et al. [9] coined the term *"Water-Scrum-Fall"* to describe a pattern which they claimed most companies implement for their software projects.

In previous research, we could confirm West's claim [10], [11]. In addition, independently conducted research [12] and a number of country-specific [4], [13], and industry-hosted studies [14] provide evidence on the use of *hybrid development methods*. However, while stating that software and system development is diverse and increasingly driven by agile methods was a first-class citizen in research, little information is available about the nature of hybrid development methods, what they look like, and how to devise them.

Problem Statement: Modern software and system development does not follow any blueprint. A variety of different frameworks, methods, and practices are used in practice; according to a study by Klünder et al. [15], 78.5% of practitioners evolve their processes over time to improve, for instance, different product quality attributes and to keep flexibility regarding the ability to react to change. However, an understanding of what a hybrid development method is composed of is missing, e.g., which combinations of frame- works, methods, and practices for software and system development help practitioners implement a process environment that provides the company and the management with a stable framework while providing developers with the demanded flexibility [6], [11].

***Objective:*** The work presented in this paper aims to lay the foundation for understanding *hybrid development methods* and to develop adaptable construction procedures that help devise such methods grounded in evidence. The objective of our research is to *understand which frameworks, methods, and practices are used to realize hybrid methods in practice and to provide an evidence-based characterization of such methods*.

***Contribution:*** Based on a large-scale international online survey, we analyze 1,467 data points that provide information about the combined use of 60 frameworks, methods, and practices. Our findings indicate that using hybrid development methods *is* the norm in modern software and system development, and that using hybrid methods happens in companies of all sizes and across all industry sectors. We identify eight base methods providing the basis for devising hybrid methods, and we statistically compute sets of practices used to embody the base methods. We contribute a statistical process that helps computing hybrid methods (including process variants) to provide advice to practitioners what (not) to include in their process portfolio.

***Context:*** The research presented in this paper emerges from the HELENA[1] study, which is a large-scale international online survey in which 75 researchers and practitioners from 25 countries participated. The study was implemented in two stages (Fig. 1) of which the first stage was a test in Europe, which was published in [10], [16]. All data and complementing materials of the second stage are available online [17].

***Outline:*** The paper is organized as follows: Section 2 presents related work. In Section 3, we present the research design. The results are presented in Section 4 and discussed in Section 5. The paper is concluded in Section 6.

## 2. RELATED WORK

A number of survey studies have sought to investigate the state of practice by focusing on software development methods. For instance, the "State of Agile" survey [14] annually collects data on the use of agile methods. The "Swiss Agile Study" [13] and the "Status Quo Agile" study [18] collect data in certain intervals aiming at observing the use of agile methods in Switzerland and Germany. Garousi et al. [4] provide an overview of the use of agile methods in Turkey. These studies explicitly focus on agile methods and cover only some of the traditional methods. Specifically, Theocharis et al. [11] provided evidence that this focus on agile is too narrow as, for instance, numerous companies and project teams remain skeptical and do not consider agile methods as the "Silver Bullet" [6], [19]–[21].

Companies develop a heterogeneous process portfolio comprised of a variety of traditional and agile methods and practices. Cockburn [22] described a framework to choose appropriate methods to address the needs of projects. Boehm and Turner [23] aimed to overcome situation-specific shortcomings of agile and plan-driven development by defining five factors that describe a project environment and help determine a balanced method. Different complementary research streams were developed to tackle process variability and adaptability demands. For instance, Clarke and O'Connor [24] provide collections of situational factors supporting process tailoring. Another research stream is focused on software process lines, e.g., [25], [26]. All these initiatives aim to bring more flexibility to processes and to help companies devise context-specific (hybrid) processes. For such combined processes, West et al. [9] coined the term *"Water-Scrum-Fall"*, and different studies, e.g., [11], [12], [27], provide evidence that the use of hybrid methods has become the norm.

In [10], [16], we initially studied the state of practice in using different frameworks, methods, and practices in combi- nation and derived process clusters that form hybrid development methods. Klünder et al. [15] studied the development of hybrid methods and found an evolutionary approach to be the common way to devise such methods, followed by planning a method as part of software process improvement programs. This study extends the available research by investigating the characteristics of hybrid development methods with a specific focus on the components hybrid methods are made of. We statistically analyze combinations of frameworks, methods, and practices to find such combinations that have a high level of agreement among the study participants and, thus, can be considered common sense about the basic structure of hybrid methods. Beyond the plain analysis, we also make a first step towards constructing hybrid methods by describing a statistical procedure that helps in computing hybrid methods and their variants from data.

## 3. RESEARCH DESIGN

We present the research design including the research questions, information about the used survey instrument, and the different procedures regarding data collection, analysis and the validity procedures. The overall research design is outlined in Fig. 1, the individual steps are described in the following paragraphs.

### A. Research Objective and Research Questions

The overall objective of the research presented in this paper is to understand which frameworks, methods, and practices are used to realize hybrid development methods in practice and to provide an evidence-based characterization of such methods. For this, we study the following research questions:

***RQ1:*** *Which frameworks and methods form the basis for devising hybrid development methods?* This question sets the scene by analyzing the more comprehensive frameworks and methods that form the basis for hybrid methods and bind the different smaller practices together. This research question is motivated by a finding from our previous study [10] that process clusters are formed around "centers". The first step is thus to identify such centers. As the HELENA study contains a flag that indicates if a specific set of frameworks, methods and practices is intentionally used in combination, the analysis is performed twice: once for the entire dataset and once for the subset of data for which the study participants explicitly stated to combine the different processes.

[1] HELENA: Hybrid dEveLopmENt Approaches in software systems development, online: https://helenastudy.wordpress.com

***RQ2:*** *Which practices are used to embody method combinations for devising hybrid methods?* Having identified the base methods and the method combinations providing the frame for a hybrid method, we analyze the data for recurring practices used to embody the identified base methods and method combinations. That is, we aim to identify specific combinations of frameworks, methods *and* practices that, together, form hybrid development methods. Again, the investigation is performed twice for the entire dataset and the subset of participants that explicitly combine processes.

***RQ3:*** *How can hybrid development methods be characterized?* This question aims to develop a procedure that helps characterizing hybrid methods by defining core practices that, together with the base methods and method combinations, provide the starting point to devise specific hybrid methods. We aim to statistically define hybrid methods through the sets of practices included, and we also aim to provide a means to define a hybrid method and its variants to help practitioners decide what to (not) include into their process portfolio. Again, the investigation is performed twice.

## B. Instrument Development and Data Collection

Data was collected using the survey method [28]. We designed an online questionnaire to solicit data from practitioners about the development approaches they use in their projects. The *unit of analysis* was either a project (ongoing or finished) or a software product.

***1) Instrument Development and Structure:*** The survey instrument was developed and refined in several iterations (Fig. 1; see [15] for further details). Finally, the research team included 75 researchers from all over the world. The questionnaire was made available in English, German, Spanish, and Portuguese and consisted of five parts (with number of questions): Demographics (10), Process Use (13), Process Use and Standards (5), Experiences (2), and Closing (8). In total, the questionnaire comprised up to 38 questions, depending on previously given answers [17].

***2) Data Collection:*** The data collection period was May to November 2017 following a *convenience sampling strategy* [28]. The survey was promoted through personal contacts of the participating researchers, posters at conferences, as well as posts to mailing lists, social media channels (Twitter, Xing, LinkedIn), professional networks, and websites (ResearchGate and researchers' (institution) home pages).

## C. Data Analysis Procedures

The data analysis consisted of multiple parts, which are described in detail in this section.

***1) Data Cleaning and Data Reduction:*** The first step was the preparation of the data. We opted for the full dataset of the second stage of the HELENA study [17], which consists of 1,467 data points. As many questions were optional and participants had the opportunity to skip mandatory questions, we first analyzed the data for *NA* and *-9* values.

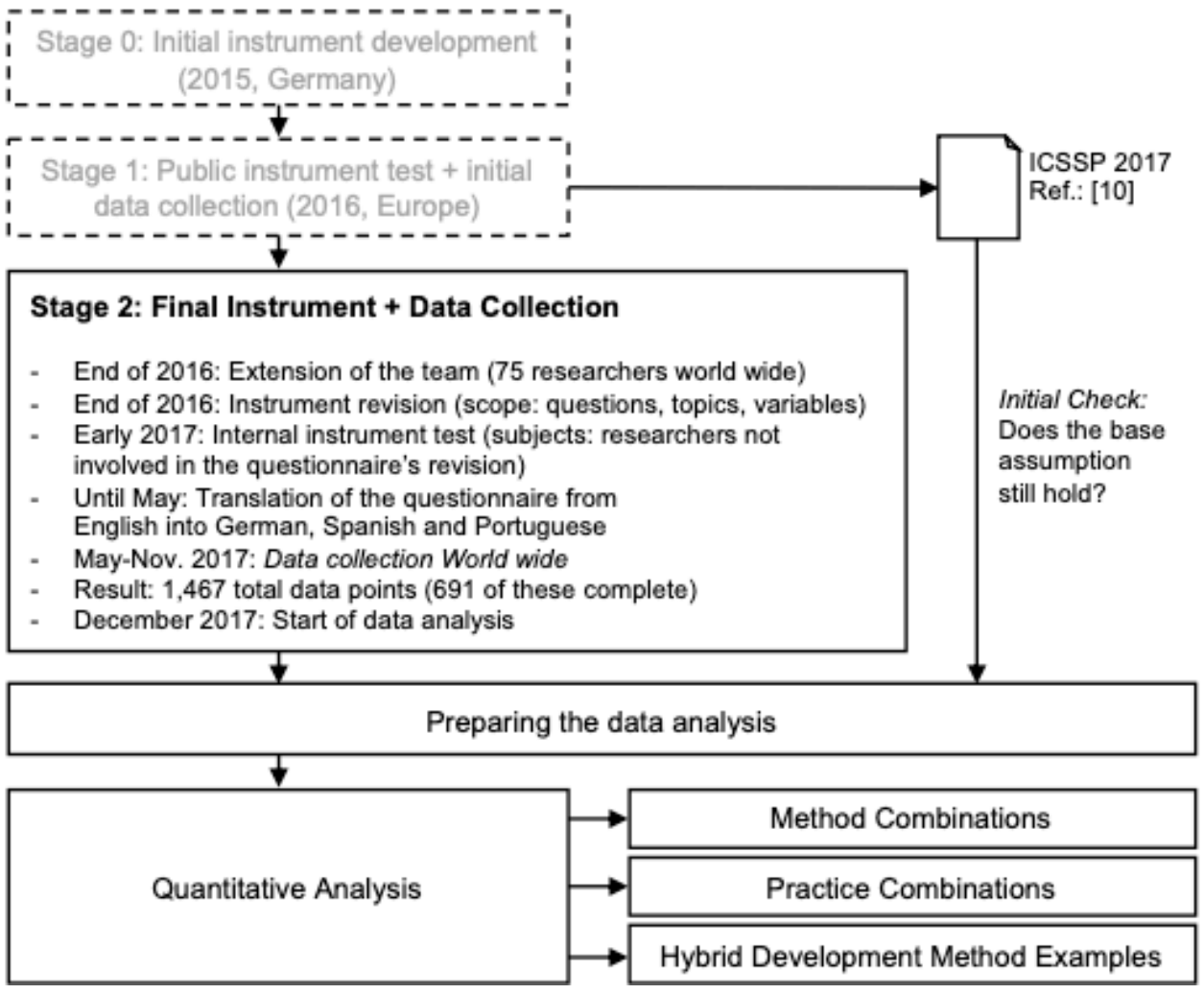

Fig. 1. Overview of the research method applied to this study including the study's position in the overall HELENA project

While *NA* values indicate that participants did not provide information for an optional question, *-9* indicates that participants skipped a mandatory question. Depending on the actual question, *-9* values were either transformed into *NA* values or the respective data point was excluded from further analysis as we considered the question not completely answered. Finally, in the question about company size (D001; [17]), we combined the categories *Micro* and *Small* into a new category *Micro and Small (1–50 employees)* leading to an almost even distribution among all company sizes.

***2) Checking the Base Assumptions****:* In this study, we are interested in the particular process combinations used in industry. Our base assumption is that frameworks, methods, and practices are combined in practice as claimed by West et al. [9]. For this, in our previous studies [10], [16], we quantitatively analyzed the initial data using a set of hypotheses. As the first step in the quantitative data analysis, we tested the two hypotheses shown in Table I using Pearson's $\chi^2$ test at a significance level of $p \leq 0.05$.

Table I: Hypotheses used to check the base assumption that combinations are common practice

| **Hypotheses** | |
|---|---|
| $H1_0$ | The use of hybrid approaches does not depend on the company size. |
| $H2_0$ | The use of hybrid approaches does not depend on the industry sector. |
| **Question/Variable Assignment to Hypotheses** | |
| $H1_0$ | Combination (PU04), Company size (D001) |
| $H2_0$ | Combination (PU04), Industry sector (D005) |

While H1 was directly tested using Pearson's $\chi^2$ test, testing H2 required a different procedure as participants were able to provide different industry sectors as targets for the question D005 [17]. Therefore, a Pearson's $\chi^2$ test was evaluated for all industry sectors. For each industry sector, we tested the share of participants stating that they (do not) combine the different frameworks, methods, and practices and compared those with all the other industry sectors. As the number of data points per industry sector influences the p-value, we used all selections of the respective industry sectors as sample size for the $\chi^2$ tests. Finally, as we tested a single hypothesis using multiple tests, we used a

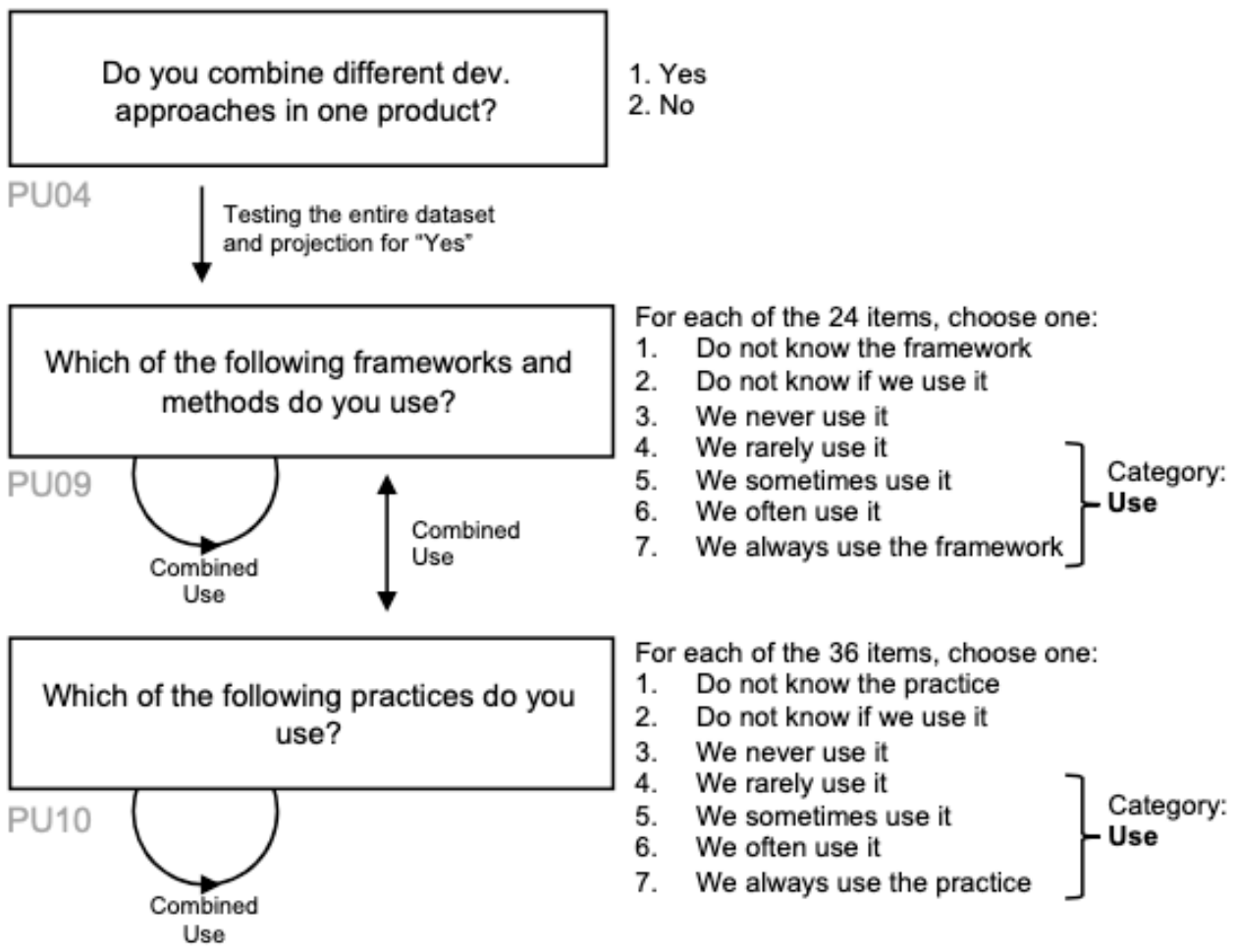

Fig. 2. Overview of the analysis model used in this study

*Bonferroni correction*[2] to adjust the significance level by dividing the given significance level of $p \leq 0.05$ by the number of tests. Including the option "Other" in question D005, we provided 20 industry sectors to choose from, i.e., the corrected significance level is $p_{B_cor} \leq 0.0025$.

***3) Quantitative Analysis for Process Combinations:*** To derive process combinations from the data, we analyzed the (combined) occurrence of frameworks, methods, and practices in the dataset. For this analysis, we used the questions PU09 (frameworks and methods), PU10 (practices), and PU04 (combined process use). To structure the analysis, we defined an *analysis model* (Fig. 2). The analysis was performed in multiple steps and each step was performed twice: (i) on the entire dataset and (ii) on a subset created from a filter using the participants' selection of question PU04. Specifically, the following main analyses[3] were performed:

***Methods:*** First, the combined use of the different frameworks and methods, e.g., *Waterfall*, *DevOps*, and *Kanban*, was analyzed, and a Top-10-like list of methods and method combinations was computed. The combinations were computed using a recurrence threshold of 35%, i.e., we included methods and combinations that were selected by at least 35% of the participants. The recurrence threshold was set to 35% as it identifies a minimal group of three frameworks and methods in the entire dataset, and a minimal group of four in the projected dataset generated through PU04="Yes".

***Practices:*** Similarly, we analyzed the practices, e.g., *Coding Standards*, *Code Reviews*, and *Release Planning*. Different to the analysis of the frameworks and methods, we used an 85% recurrence threshold as this threshold provides a minimal group of two practices in the entire dataset as well as in the projected dataset generated through PU04="Yes".

### D. Validity Procedures

To improve the validity and to mitigate risks, we implemented different measures focused around replicability and consistency as well as bias. First, our research is grounded in previously conducted studies. Notably, the key question of this study was derived from the outcomes of our previously conducted study [10]. An extended design team developed the survey instrument as described in [17]. The data analysis was performed by different teams, i.e., one team performed the hypothesis testing while another team focused on the quantitative analyses. Researchers not involved in the data analysis were tasked to provide the quality assurance.

Table II: Result of testing H1 (independence of company size)

| Id | Results | Decision | [10] |
|---|---|---|---|
| $H1_0$ | $\chi^2 = 1.9972$, df = 3, $p = 0.573$ | no support | no support |

Table III: results for testing H2 (independence of industry sector); corrected significance level is $p_{B_COR} \leq 0.0025$

| Industry Sector | This NH:H | Rest NH:H | $\chi^2$ | $p$-value |
|---|---|---|---|---|
| Automotive SW / Systems | 14 : 66 | 163 : 515 | 1.37 | 0.24 |
| Aviation | 09 : 21 | 168 : 560 | 0.43 | 0.51 |
| Cloud Appl. and Services | 29 : 95 | 148 : 486 | 0.00 | 1.00 |
| Defense Systems | 02 : 26 | 175 : 555 | 3.38 | 0.07 |
| Energy | 07 : 30 | 170 : 551 | 0.21 | 0.65 |
| Financial Services | 34 : 149 | 143 : 432 | 2.73 | 0.10 |
| Games | 01 : 17 | 176 : 564 | 2.32 | 0.13 |
| Home Auto. / Smart Build. | 05 : 17 | 172 : 564 | 0.00 | 1.00 |
| Logistics / Transportation | 11 : 43 | 166 : 538 | 0.14 | 0.71 |
| Media and Entertainment | 06 : 25 | 171 : 556 | 0.10 | 0.75 |
| Med. Devices / Health Care | 13 : 61 | 164 : 520 | 1.20 | 0.27 |
| Mobile Applications | 19 : 105 | 158 : 476 | 4.82 | 0.03 |
| Other Emb. Systems / Svcs. | 09 : 46 | 168 : 535 | 1.22 | 0.27 |
| Other Information Systems | 20 : 87 | 157 : 494 | 1.22 | 0.27 |
| Public Sector / Contracting | 21 : 72 | 156 : 509 | 0.00 | 0.95 |
| Robotics | 01 : 17 | 176 : 564 | 2.32 | 0.13 |
| Space Systems | 08 : 26 | 169 : 555 | 0.00 | 1.00 |
| Telecommunication | 07 : 38 | 170 : 543 | 1.19 | 0.27 |
| Web Appl. and Services | 40 : 162 | 137 : 419 | 1.68 | 0.20 |
| Other | 27 : 63 | 150 : 518 | 2.12 | 0.15 |

Second, as one of the main goals of this study is to build a quantitative basis, we opted for the *convenience sampling strategy* [28] to collect the data by accepting the risk of losing full control in terms of sampling, response rate and so forth. This decision was made to collect as many data points as possible. To handle this risk, before analyzing the data, we implement rigorous data pre-processing including a consistency check of the data (see Section 3-C1).

## 4. RESULTS

We present the results following the structure of the research questions (Section 3-A) and our analysis model (Fig. 2).

### A. Checking the Base Assumptions

As outlined in Section 3-C2, our study is built on previously published studies [10], [16] that found no evidence that the use of combined processes in practice depends on company size or industry sector. Therefore, we tested two hypotheses for which the results are presented in Table II (H1; from [10]) and in Table III (H2; according to [16]). Table III

[2] The Bonferroni correction is used when several statistical tests are performed simultaneously. This requires an adjustment of the α value [29].

[3] The analysis was performed using the R-package apriori (online: https://www.rdocumentation.org/packages/arules/versions/1.6-2/topics/apriori) that, among other features, allows for setting recurrence thresholds and mini- mum/maximum set sizes.

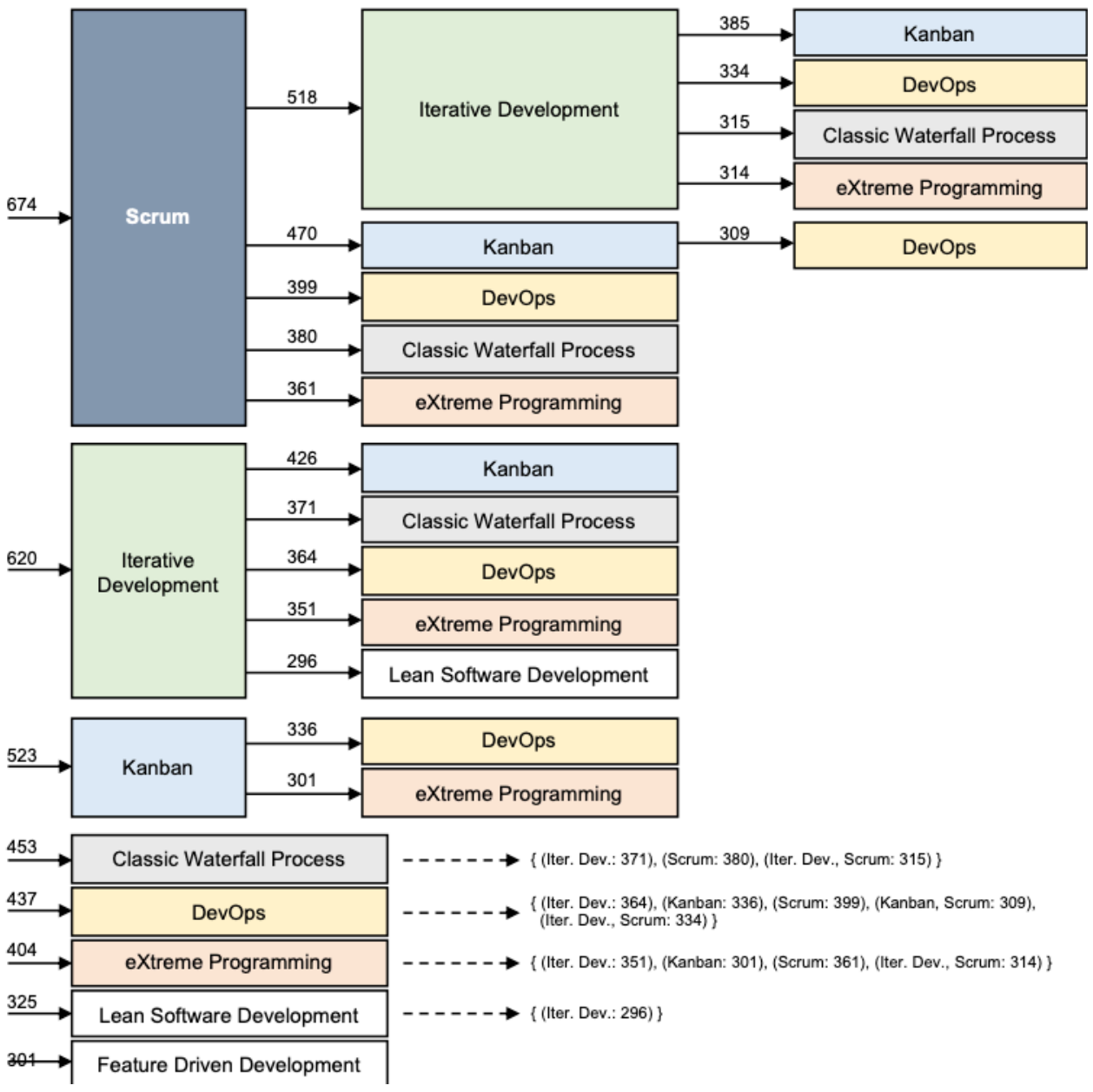

**Fig. 3.** Base methods and method combinations (35% threshold) in the non- filtered dataset. The figure reads from the right to the left, e.g., 309 participants use Scrum, Kanban and DevOps in combination

shows the ratios of participants that do not combine (NH) and those that combine processes (H) within an industry sector and for all remaining industry sectors. The table shows the individual test results, which, however, have to be considered in the context of the Bonferroni-corrected significance level of $p_{B_cor} \leq 0.0025$. The results shown in Table II and in Table III support the findings from [10], [16]. Notably, the results from Table III, given the Bonferroni correction, show that no $\chi 2$ test is significant, which does not allow for concluding that the industry sector influences the use of hybrid methods. Hence, the results show that the combined use of different frameworks, methods, and practices, i.e., the use of hybrid methods, is a common practice in industry. The leading question of this study—*what do such combinations look like*—has therefore to be considered of high relevance.

> **Finding 1:** The use of hybrid development methods has not shown any dependence with regards to either the company size (H1) or the industry sector (H2). Therefore, given the high p-value of the majority of the tests, the use of hybrid development methods can be considered state of practice across companies of all sizes and in all industry sectors.

## B. Combined Use of Frameworks and Methods

The first step in the quantitative analysis is the investigation of the combined use of frameworks and methods. Of the 1,467 data points, 845 provide data for question PU09 (Fig. 2). As shown in Fig. 2, this multiple-choice question provided 24 items to choose from complemented with a free-text option. Of the 845 data points, 792 had multiple selections.

Figure 3 shows the resulting combinations using the 35% threshold for the combined process use in the entire (non-filtered) dataset. This threshold results in 17 groups of two or three combined frameworks and methods—there is no group with four or more elements with at least 35% agreement regarding the combined process use. *Scrum* is the most frequently selected method (674 participants),

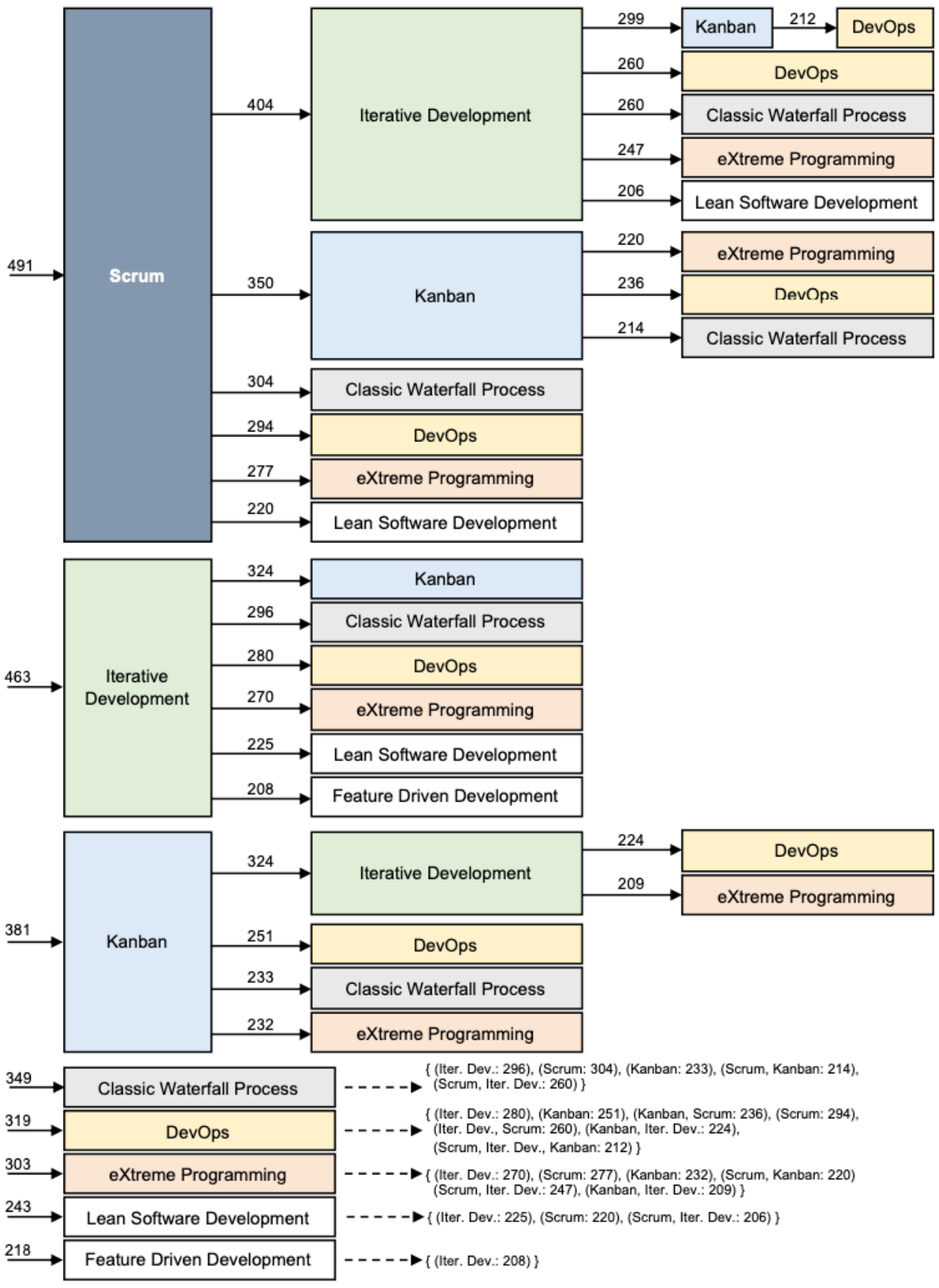

**Fig. 4.** Refined base methods and method combinations (see Fig. 3; 35% threshold) in dataset filtered for question PU04.

which is followed by *Iterative Development* (620) and *Kanban* (523). Extending the scope to framework and method combinations, a number of couples and *all* triplets include *Scrum*. Expected combinations are present, e.g.,

*(Scrum–Kanban–DevOps)* (309 participants), and the "Water-Scrum-Fall" ([9]; 380 participants).

Applying question PU04 (Fig. 2) as a filter, i.e., re-running the analysis for only those participants that explicitly claimed to use the different frameworks, methods, and practices in combination, Fig. 4 results in 27 groups of two to four combined frameworks and methods, whereas there is no group with five elements or more having at least 35% agreement regrading combined process use. The combined frameworks and methods as shown in Fig. 3 and in Fig. 4 *do not* provide the full picture as they only form the "core", but are complemented with further frameworks, methods, and practices, which will be elaborated in more detail in the following sections.

> **Finding 2:** Among the 24 frameworks and methods presented to the study participants, we identified 17 (entire dataset, Fig. 3) and 27 (dataset filtered for question PU04, Fig. 3) core groups with two to four elements for which the study participants agree with at least 35% on their combined use. These combinations are based on eight *base methods* that provide the frame for hybrid development methods.

## C. Combined Use of Frameworks, Methods, and Practices

The second step in the quantitative analysis is the investigation of the combined use of frameworks, methods *and* practices (Fig. 2, PU09 and PU10). Of the 1,467 data points, 769 provide data. As shown in Fig. 2, this multiple-choice question provided 36 items to choose from and a free-text option. Of the 769 data points, 742 had multiple selections.

As described in Section 3-C3, to analyze the combinations of practices within the base methods and method combinations provided in Section 4-B, we used an 85% threshold for the agreement regarding combined use. That is, for each method combination identified in Section 4-B, the combinations of practices within these have been computed. All analyses were performed using base methods and method combinations resulting from the entire dataset and from the projected dataset based on the answers to the question PU04 (Fig. 2). The overall result is shown in Fig. 6, which will be described step by step in the following paragraphs.

***1) Unfiltered Practices**:* As a first step, the (non-filtered) dataset was analyzed for the most commonly used practices, i.e., those practices with the highest agreement regarding combined use without a particular combination of methods. To find these practice combinations and to find those groups that have the largest agreement in the entire dataset, we explored the dataset. The smallest group with the highest agreement in the data was the pair *Code Review* and *Coding Standards* (n=674, agreement=0.87). The agreement level of 0.87 was also used to set the threshold of 85% agreement as introduced in Section 3-C3.

The results of the analyses of the entire (non-filtered) dataset using the 85% threshold are shown in Fig. 5. The figure shows for the entire dataset three practices (*Code Review*, *Coding Standards* and *Release Planning*) for which one pair of two practices has an 85% agreement. Likewise, in the projected dataset (after applying PU04 as filter), five practices could be identified (*Code Review*, *Coding Standards*, *Release Planning*, *Automated Unit Testing* and *Protoyping*). Of these five practices, three pairs of two composed from the five practices could be identified, which have at least 85% agreement among the participants of the study.

***2) Individual Practices:*** In the same reading as for Fig. 5, the upper part of Fig. 6 presents the practices reaching 85% agreement within the context of the respective base methods and method combinations. The upper-left part of Fig. 6 presents the results for the entire dataset, while the upper- right part of Fig. 6 presents the results for those base methods and method combinations computed from the projected dataset after applying PU04 as a filter (Section 4-B).

For each practice (Fig. 2; PU10, 36 items), Fig. 6 shows the assignment to a base method or a method combination for which 85% agreement could be found in the dataset. The total number of such practices assigned to a particular method combination is shown in the row "Number of practices in combinations" beneath the respective method combinations. For instance, for the method combination *(Scrum–Kanban)*, 14 practices are assigned to this method combination in the entire dataset and, respectively, 15 practices are assigned to this method combination in the PU04-projected dataset. All possible combinations of frameworks, methods, and practices with 85% agreement are constructed from these individual practices, which is elaborated in more detail in Section 4-C3.

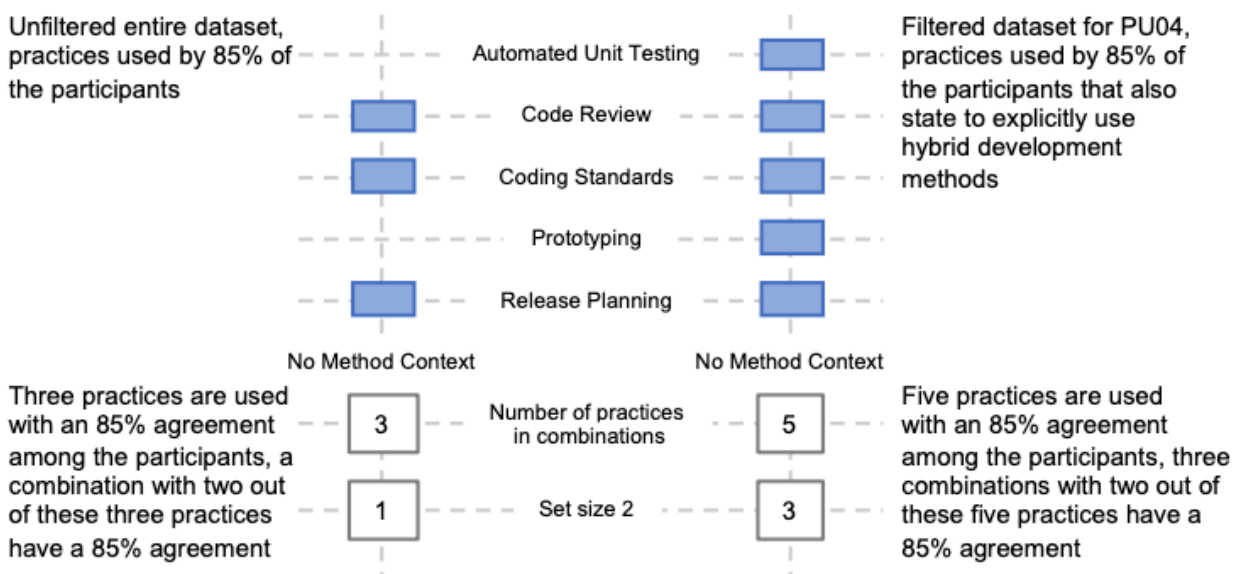

**Fig. 5.** Overview of the most frequently used practices in the entire dataset (left) and in the filtered dataset (PU04, right). The figure illustrates the most frequently used practices and also shows how many possible combinations can be found with 85% agreement among the participants.

The visualization in the upper part of Fig. 6 allows for two main observations: first, the sparsity of rows and thus a limited number of practices consistently selected by the participants in the context of a given method or method combination. Second, the selected practices (highlighted rows) are consistent across different method combinations. That is, a limited number of practices is consistently used with an agreement of at least 85% regardless of the actual method combination. In addition, two minor observations can be made: first, the "density" of the practices for which the participants agree regarding their combined use is higher in the PU04-projected dataset than in the non-filtered dataset, i.e., in that share of the data in which the participants explicitly stated to combine multiple frameworks, methods, and practices. Second, it seems that as if the larger the number of combined methods is the more practices find an agreement among the participants. For instance, the rightmost method combination in Fig. 6 *(Scrum–Iterative Development–Kanban–DevOps)* has 21 practices assigned for which the participants find an agreement of at least 85%.

***3) Combinations of Practices:*** The lower part of Fig. 6 extends the analysis from Section 4-C2 starting with the "Number of practices in combinations" row. This row shows how many practices are assigned to the different method combinations thus forming the basis for framework, method, and practice combinations to derive hybrid development methods. Within these sets of practices, we search for *practice tuples* of increasing size having agreement of at least 85% and that are used in combination in the respective method combination. Taking the combination *(Scrum–Kanban)* as an example, 14 (entire dataset) and 15 (PU04-projected dataset) practices are assigned to this combination. In the first step, we search for pairs of two practices from these 14 (15) practices with the required agreement level. This results in 48 pairs (for the entire dataset) and, respectively, 65 pairs (PU04-projected dataset) of practices out of 14 (15) practices. In the next step, we search for 3-tuples, then for 4-tuples, and so forth until no *x*- tuple with the required agreement level is found. As Fig. 6 shows, the biggest set size with an agreement of at least 85% is eight. For instance, in the PU04-projected dataset and for the combination *(Scrum–Iterative Development–Kanban–DevOps)*, three 8-tuples of practices from the 21 practices assigned to this combination can be found in the dataset.

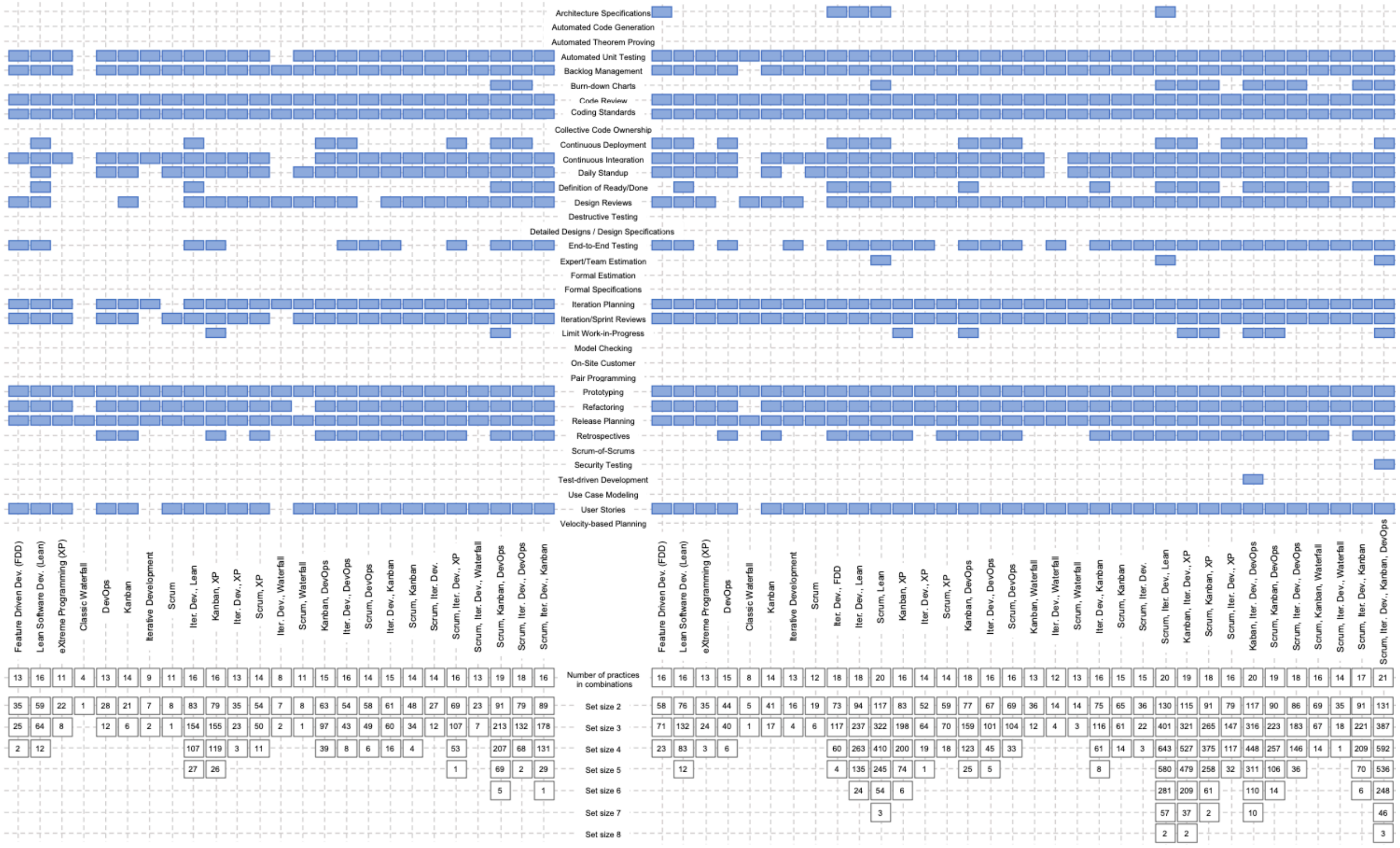

Fig. 6. Overview of the practices used with 85% agreement in hybrid development methods. The left part of the figure shows the analysis results for the entire, non-filtered dataset, the right part illustrates the analysis results for the filtered dataset based on the value of question PU04. The upper part of the figure shows the practices used together with the different method combinations, and the lower part shows the of possible combinations of practices of given set sizes within the different method combinations, based on the *number of practices in combinations*-row.

The lower part of Fig. 6 allows for two main observations: first, the larger the combination size the more agreement can be found regarding the practices that would be included in a method or method combination. Second, the larger the number of combinations within a group the more practices are consistently selected by the participants. Also, similar to the observations from Section 4-C2, we see that the larger the number of combined methods, the more combinations of practices find agreement among the practitioners, and the size and the amount of combinations are bigger in the PU04- projected data than in the entire dataset.

**Finding 3:** Analyzing the 36 practices and their relation to the methods and method combinations found in Section 4-B, we find few practices only that find agreement of at least 85% among practitioners. However, as shown in Fig. 6 the assignments show a consistency across the base methods and method combinations. That is, few practices only are consistently used for hybrid development methods.

### D. Constructing Hybrid Development Methods

The analyses conducted in Section 4-B and Section 4-C provide important insights regarding the *base methods*, the basic *method combinations* and the *number of practices assigned* to these base methods and method combinations. In this section, we demonstrate how to use our analysis method to incrementally construct hybrid development methods. For this, we apply the following procedure:

1. Based on the smallest groups of practices for the entire dataset and for the PU04-projected dataset, which are shown in Fig. 5, we form the “core” of practices from the different pairs: one pair for the entire dataset and three for the PU04-projected dataset.
2. For each base method or method combination, we add the core(s) to set an extended method context. For instance, instead of looking for all practices to be combined with *Scrum*, we search all additional practices meeting the required agreement level of 85% for the new combination *(Scrum–core*i*)* with i denoting the cores identified.
3. In the final step, we integrate all frameworks, methods, and practices into hybrid development methods by building the unique combinations (the process variants) of all these components.

That is, we aim to identify those practices (if any) that are included in bigger combinations containing the “core” for each of the base methods or method combinations. We applied this procedure to both the entire dataset and the PU04- projected dataset, which results in the *(method-core*i*-practice)* combinations shown in Fig. 7 for the entire dataset and Fig. 8 for the PU04-projected dataset.

Taking the “plain” *Waterfall* as an example, we see in both figures that the *Waterfall* is characterized by the core-practices only. Moving on to “Water-Scrum-Fall”, i.e., the combination *(Scrum–Waterfall)*, in the entire dataset (Fig. 7), we see one single combination of size three containing the *(Scrum–Waterfall)* method combination, the core consisting of *(Code Review–Coding Standards)* and *Release Planning* as third practice. In the PU04-projected data (Fig. 8), "Water- Scrum-Fall" is characterized by the *(Scrum–Waterfall)* method combination and three cores of which either is extended by (*Prototyping–Iteration/Sprint Reviews*).

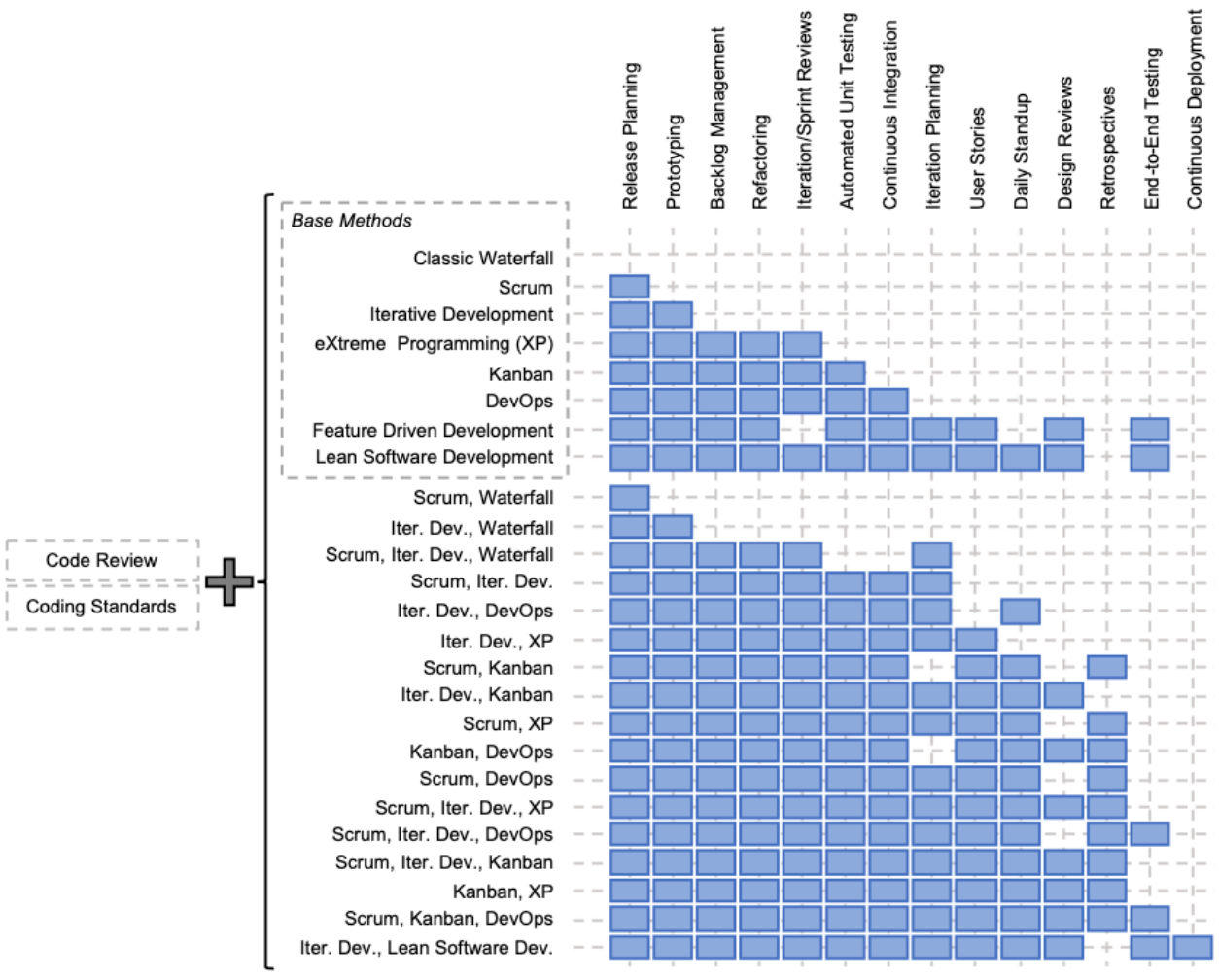

Fig. 7. Characterization of base methods (top) and method combinations (bottom) generated by identifying recurring combinations of three practices with at least 85% agreement level that contain the core of the entire non- filtered dataset, i.e., (*Code Review–Coding Standards*).

Both figures also show the gradual increase of the practice pools. Staying with *Scrum* in the entire dataset (Fig. 7), i.e., (*Scrum–Code Review–Coding Standards–Release Planning*), the base method *Iterative Development* "extends" this combination with *Prototyping*, i.e., the combination would be (*Iterative Development–Code Review–Coding Standards– Release Planning–Protyping*). That is, *Scrum* is characterized by the practice combination (*Code Review–Coding Standards– Release Planning*) and *Iterative Development* is characterized by (*Code Review–Coding Standards–Release Planning– Prototyping*).

**Finding 4:** Applying our analysis procedures, we can define a statistical construction procedure for describing hybrid development methods. Also, our description is not only limited to specific hybrid method instances; we can also characterize a hybrid method *and* its process variants.

# 5. DISCUSSION

Having presented our results, we conclude this paper by answering the research questions, discussing our findings and discussing the threats to validity.

## A. Answering the Research Questions

The findings provide a rich quantitative basis and evidence to answer our research questions posed in Section 3-A: RQ1: Klünder et al. [15] found that processes mainly *evolve* into hybrid methods, and provided evidence and generalized the claim by West et al. [9] about the *"Water-Scrum-Fall"*. With this paper, we provide insights regarding eight base methods that are recurrently combined to form hybrid development methods (cf. Finding 2).

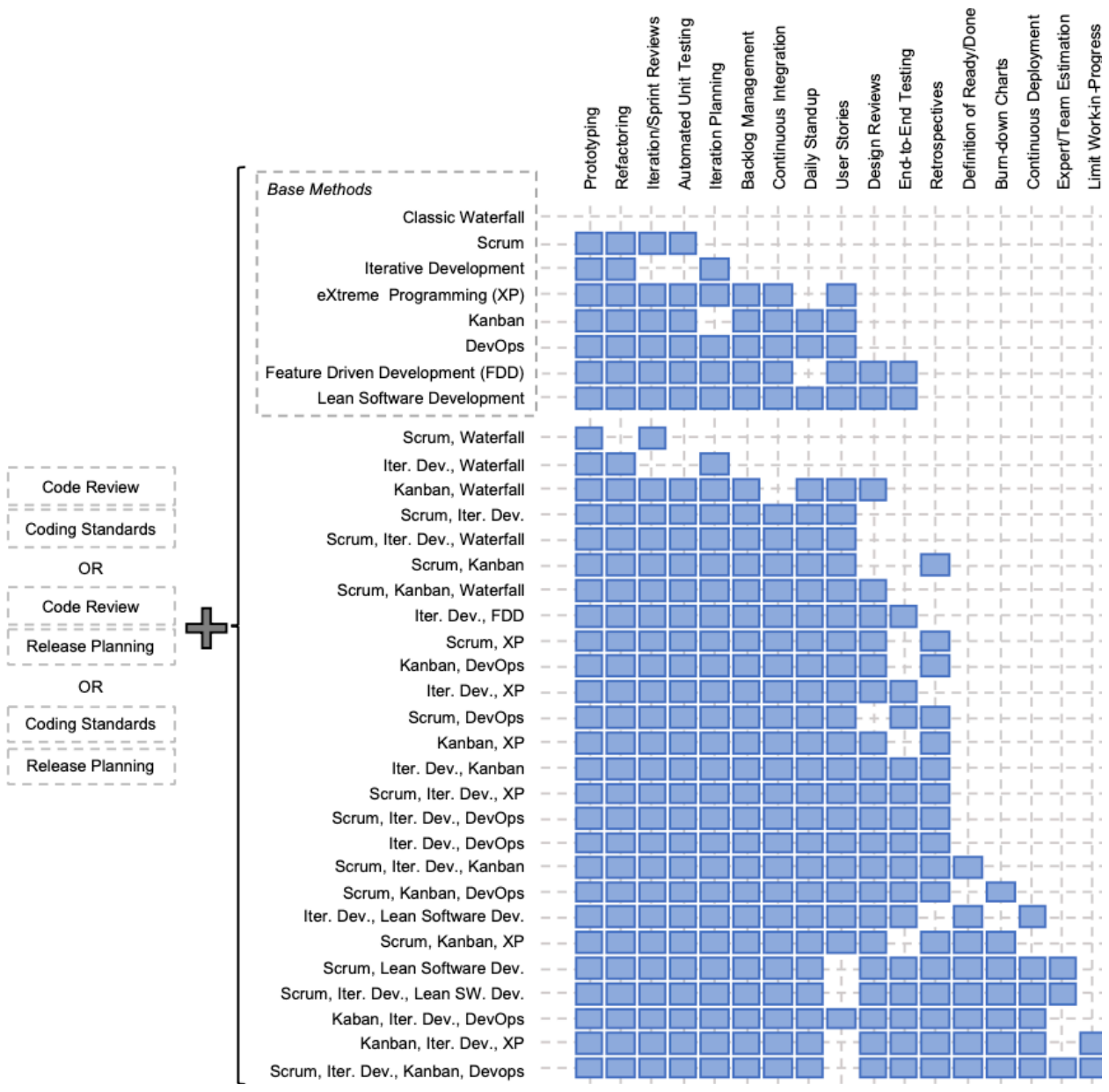

Fig. 8. Refined characterization using the dataset filtered for question PU04 of base methods (top) and method combinations (bottom) (see Fig. 7). The three cores used are: (*Code Review–Coding Standards*), (*Coding Standards– Release Planning*), (*Code Review–Release Planning*).

**RQ2:** In this paper, we identify the most frequently used practices and how these are combined with each other. Our results reveal a small *core* of practices used by practitioners regardless of the (hybrid) development method selected (cf. Finding 3).

**RQ3:** In modern software and system development, methods and practices are often combined into processes that are context-dependent. However, when attempting to characterize the different methods by systematically constructing the set of practices using a bottom-up strategy, we show that the resulting combinations of practices vary very little and consistently repeat the same practices (cf. Finding 4).

## B. Hyped Methods and Old Practices

So far, we could identify eight base methods and few practices that, together, are the heart of hybrid development methods. Figure 7 and Fig. 8 show these key components that find agreement across the participants of the HELENA study. Looking closer at the practices, we see that hybrid methods are heavily composed of (mostly technical) practices that have been used in software development for decades, namely *Code Review*, *Coding Standards* and ***Release Planning.*** On the one hand, the sets of practices and their assignment to methods shows obvious similarities (Fig. 6). On the other hand, it is not trivial to identify a specific method or method combination from the practices used in a specific context.

Accepting that formally defined methods are not applied in practice ([30]–[32]) and that hybrid development methods are the norm ([9]–[11] and Section 4-A), we can pose a number of further questions. From our perspective, the most urgent question is for the actual role of methods, for instance: What is the value of stating that someone, e.g., uses Scrum, when no-one does it by the book? What are the implications for devising a particular method if, in practice, the method starts evolving [15] into a hybrid? What are the implications for software process improvement programs if, regardless of the "showroom method" providing the umbrella, the practices are the *stable* key components of organizing and conducting the actual project work? What are the implications for educators if companies require students trained in latest methods, but key is a solid understanding of the "old stuff", which is still the core of modern software engineering practice?

Brooks [33] argued that there is no *"Silver Bullet"* that would fit all the different flavors that the software engineering industry has. However, especially in the last decade, it ap- pears that Brooks' observation has been forgotten, and some methods are relentlessly advertised as *the* silver bullet. *New* methods are continuously spawned, and people engage in rather unhealthy discussions arguing whether one method is superior over another. So, are we chasing the white rabbit by focussing on hyped methods and substituting agility with Scrum? One could disagree as "revolutions" like the Agile Manifesto changed the industry inside-out, which is certainly true and industry did progress. Yet, such revolutions changed the mindset and the culture of companies, *not* the practices. Test-driven development, continuous integration, continuous delivery, continuous deployment are incarnations of building blocks that already existed, but, in Kent Beck's words[4], have been "cranked up all the knobs to 10". What has changed is how they are combined and how they are used. We argue that we should distance ourselves from such discussions about the "right" method, but should focus our attention to the practices. Studying the nuances behind practices and their implementation in different contexts would possibly lead to interesting findings whether some hybrid methods (sets of practices) are more effective than others.

As shown in this paper, among practitioners, strong agreement can be found at the practice level, and, when analyzing at this level of granularity, methods and frameworks step into the background. Therefore, we argue that *researchers* should report on the actual practices when presenting cases, as assumptions on practices used based on a method or framework do not hold, *practitioners* should be mindful about new hypes as the identified core practices are building blocks that are agnostic of methods and, finally, *educators* should put more emphasis on teaching practices rather then methods by explaining the rationale behind them and the different ways in which they can be executed.

### C. Threats to Validity

We discuss the threats to validity according to [34]. ***Construct Validity:*** As we used an online questionnaire, the options provided in the multiple-choice questions might have been incomplete, and that participants misunderstood some questions, which probably led to incomplete or wrong answers. To reduce these threats, multiple-choice questions were complemented with a free-text option, and a team of researchers constructed the questionnaire, tested and revised it accordingly (Section 3-B). The questionnaire was published in four languages to reduce the risk of misunderstandings due to language issues. The convenience sampling strategy could lead to participants not fully reflecting the target audience. As executing the survey required specific knowledge and a qualitative analysis of the free-text answers in [15] only resulted in meaningful answers, we consider this threat to be mitigated. *Internal Validity:* Prior to the analysis, we cleaned the data (Section 3-C1), which could introduce a threat. Also, in the data analysis, we did not exclude data points per se, but performed the analyses with varying n's. To mitigate the risks, all steps have been per- formed by at least two researchers and have been checked by other researchers not involved in the actual analysis activities. ***Conclusion Validity:*** The interpretation of the statistical tests is based on a significance level of $p \leq 0.05$, and we found no evidence that allows us to reject our null hypotheses (Table I). Furthermore, for analyzing sets of methods and practices, we used a 35% and an 85% threshold (Section 3-C3). Changing these thresholds would influence the results by enlarging the sets of methods and practices. Also, the limited set of options for the multiple-choice questions could influence the findings. Extra research is necessary to study the effects in more detail. ***External Validity:*** Although our analysis is based on a large dataset (Section 3-C1), we cannot claim full generalizability. Parts of the dataset provide equal distributions such that we can draw conclusions, e.g., for the company size or the industry sector. For other factors, further research is necessary.

## 6. CONCLUSION

In this paper, using a large-scale international online survey, we studied the use of *hybrid development methods* in practice. An analysis of 1,467 data points revealed that using different frameworks, methods and practices in combination as hybrid methods is the norm across companies of all sizes and industry sectors. We identified eight base methods and few practices only that find agreement among study participants. For the study participants that explicitly stated to use processes in combination, we could identify 27 base methods and method combinations that, together with three practices forming three pairs, build the basis to devise hybrid methods. We also found that the sets of practices have limited dependencies to the methods. We therefore argue that practices are the building blocks for devising hybrid methods.

In terms of future research, we plan to build on our observations and findings showing that practices are the essential unit of analysis when looking at software development activities within an organization. We note the core set of practices along with the complementary sets of practices identified in Section 4-D are common to all development methodologies. Because they are so widely deployed, we observe that development organizations see these practices as essential activities enabling them to deliver good software to their customers. We believe that the idea of having a set of common practices that are essential to sound software development has been the motivation behind maturity model frameworks like the CMMI, ISO/IEC 15504 and others. For our future work, we would like to conduct further analysis using the HELENA dataset to explore what having a core set of practices means regarding how industry views the value of maturity model frameworks and specific key process areas within those frameworks.

[4] Taken from an interview by informIT, March 23, 2001: http://www.informit.com/articles/article.aspx?p=20972, last access: February 6, 2019.

## ACKNOWLEDGMENTS

We thank all the study participants and the researchers involved in the HELENA project for their great effort in collecting data. *Dietmar Pfahl* was supported by the institutional research grant IUT20-55 of the Estonian Research Council as well as the Estonian IT Center of Excellence (EXCITE).